\documentclass[prb,twocolumn,showpacs,epsfig,epsf]{revtex4}
\usepackage{amsmath,amsthm,amsfonts,amscd,mathrsfs,pifont,bm,dsfont}
\usepackage{eucal,graphicx,color}
\renewcommand{\atop}[2]{\genfrac{}{}{0pt}{}{#1}{#2}}

\newcommand{\abs}[1]{\left\vert#1\right\vert}

\renewcommand{\i}{\mathrm{i}}
\newcommand{\tr}{\textrm{Tr}\;}

\newcommand{\1}{\mathds{1}}

\begin{document}
\title[Accumulation effects in modulation spectroscopy]{Accumulation effects in modulation  spectroscopy with high \\ repetition rate pulses: recursive solution of optical Bloch equations.}
\author{Vladimir Al. Osipov}
\email{Vladimir.Al.Osipov@gmail.com}
\author{T\~onu Pullerits}\affiliation{Chemical Physics, Lund University, Getingev\"agen 60, 222 41, Lund, Sweden}

\begin{abstract}
Application of the phase modulated pulsed light for advance spectroscopic measurements is the area of growing interest. The phase modulation of the light causes modulation of the signal.  Separation of the spectral components of the modulations allows to distinguish the contributions of various interaction pathways. The lasers with high repetition rate used in such experiments can lead to appearance of the accumulation effects, which become especially pronounced in systems with long-living excited states. Recently 
it was shown, that such accumulation effects can be used to evaluate parameters of the dynamical processes in the material. In this work we demonstrate that the accumulation effects are also important in the quantum characteristics measurements provided by modulation spectroscopy. In particular, we consider a model of quantum  two-level system driven by a train of phase-modulated light pulses, organised in analogy with the 2D spectroscopy experiments. We evaluate the harmonics' amplitudes in the fluorescent signal and calculate corrections appearing from the accumulation effects. We show that the corrections can be significant and have to be taken into account at analysis of experimental data. 
\end{abstract}
\date {\today}
\maketitle
\setlength{\textfloatsep}{20pt plus 4pt minus 4pt}
\setlength{\floatsep}{20pt plus 4pt minus 4pt}
\setlength{\columnsep}{9pt}

{\bf Introduction.}
High repetition rate, tens of MHz, laser systems produce trains of short pulses at relatively low energy. Still, because the pulse can be very short ($<$ 10 fs), the field strength during the pulses is sufficient for inducing nonlinear optical effects~\cite{BoydBook}. In addition, if the studied material does not reach the ground state equilibrium during the time interval between the pulses ($\sim$ 10 ns) accumulation effects can occur. In recent years, application of the phase modulated pulsed light for the purposes of advanced spectroscopic measurements became an area of growing interest~\cite{our, KWSMLPM2014, LBSE2017, KKMP2016, BBS2015, TDM2003, TLM2007, BBS2017, M2016}. In this technique the phase modulation of light pulses generates harmonics in the measured signal allowing extraction of information about the various light-induced coherent and dissipative dynamics. For the systems with long characteristic life-time of the excited states the accumulation effects may become essential. Accounting of such effect as well as the question of new information, which can be obtained from it, is an important theoretical and practical problem. 

The accumulation effects occurring due to irradiation of matter by the impulsive lasers can have several manifestations, such as heat accumulation~\cite{TPV2009} or particle shielding~\cite{NJSH2014}. In this article we discuss somewhat more delicate experiments addressing the spectroscopic measurements of matter characteristics. In the latter context the accumulation effects were discussed in Ref.~\cite{our}, where dynamics of the system was modelled by a (classical) rate equations. There the effects were described theoretically and verified experimentally. In the experiment the light pulses are organised in two collinear beams and arrive to the sample at the same time. The laser frequency is chosen in such a way that the system is excited by the two-photon absorption. The base optical frequencies of the two beams are modulated by acoustic frequencies $\phi_1$ and $\phi_2$, respectively. Due to the non-linear absorption, the effective intensity, $I_n$, of the light absorbed from the $n$th pair of pulses oscillates with $n$ on the frequencies $\phi=\abs{\phi_1-\phi_2}$ and  $2\phi$, it is $I_n\propto  \cos( \phi t_0 n)+ \frac{1}{4} \cos( 2\phi t_0 n)$, where $t_0$ is the time-interval between the pulses (in practice $t_0\sim$10ns, which is around $10^3$ times less then the modulation period, $2\pi/\phi$). As the result, the population of the system excited state oscillates in time as well and generates, in turn, an oscillating signal. It can be seen from the following consideration. The signal (fluorescence) is proportional to the population of the excited state $P(t)$, i.e. fraction of the total number of molecules that have been excited. As it has been shown in~\cite{our} 
the Fourier transform of the signal is given by a series of integrals. Each integral is the Fourier transform of the population between the neighbouring pulses, 
{\small \begin{equation}
\mathcal{F}\left[P(t)\right](\nu)=\sum_{n=-\infty}^\infty e^{\i \nu t_0 n}\int_{0}^{t_0} P(t-t_0 n| P_n)e^{-\i \nu t}dt.
\end{equation}
}
Behaviour of the population $P(t-t_0 n| P_n)$ after the $n$th pulse parametrically depends on the population $P_n$, which is the population taken at the instance of time right before the $n$th pulse.
One can show, that $P_n$ satisfies a non-linear recurrence 
\begin{equation}\label{recPn}
P_{n}=\kappa(P_{n-1},I_{n};K).
\end{equation}
The function $\kappa$ is structurally determined by the type of relaxation kinetics $K$ and includes the effective intensity, $I_n$ absorbed by the system from the $n$th pulse. Indeed, as soon as the population of the excited state taken right before the $n$th pulse is $P_{n-1}$, the light pulse can excite $(1-P_{n-1})$ fraction of molecules only, so that the initial condition for solution of kinetic equations, denoted by letter $K$, is $P_{n-1}$, while $I_n$ enters as a parameter. Being solved in the time-interval starting right before the $n$th pulse and ending right before the $(n+1)$th pulse the kinetic equations give the value $P_n$. The dependence of $P_n$ on $K$, $P_{n-1}$ and $I_n$ is encoded in the formula~(\ref{recPn}). For large $n$ the solution of the above recurrence~(\ref{recPn}) does not  depend on the way the system was prepared, this solution determines the "dynamical" steady-state of the system. Obviously, in case of fast decay of the excited state, meaning that $P_{n-1}=0$, the population as well as the generated fluorescence oscillates according to $I_n$. Therefore, the harmonics appearing in the signal are determined by $I_n$ only. Detection of the harmonics and their relative amplitudes can give information on the light-matter interaction pathways, while periodic repetition of the process allows to improve the signal to noise ratio for the harmonics' amplitudes, i.e. amplitudes of the peaks in the frequency domain $\nu$ of the signal. 

Below, we call the excited state generated by the laser pulses with a given $n$ a primary state until the next $(n+1)$th pulse comes. The part of the primary excited state, which remains even after the $(n+1)$th pulse, we call accumulated state. The modulated frequencies of the signal component due to the primary and accumulated excited states follow different peaks.
It was proposed in Ref.~\cite{our}, that the cumulative effects can be taken into account, but also allow to extract information of the dynamical processes in the material. Indeed, when the time interval between pulses is short and the system does not relax to its equilibrium completely, the population $P_{n-1}\ne 0$, the non-linearity of function $\kappa$ causes generation of new harmonics and bring changes into the relative amplitudes of the original harmonics (harmonics which are contained in $I_n$). Thus, measuring of the harmonics' relative amplitudes in the signal gives us information on $K$, as far as other experimental parameters are known. In particular, it was shown, that comparison of the linear kinetics and the one with quadratic term, which are causing the spontaneous decay in a molecular system ($\dot P(t)=-\Gamma P(t)$, where $P(t)$ is the population of the excited state) and the band-to-band annihilation of hole-electron pairs in a semicoductor ($\dot P(t)=-g P^2(t)$, where $P(t)$ is the concentration of free carriers) respectively, have a well defined distinct signatures. Thus, the idea to utilise the frequency modulated laser pulses for optical measurements of dynamical processes in various media seems promising, while accounting of the emerging cumulative effects requires revision of our theoretical models.

In the present work we extend the theoretical approach developed in Ref.~\cite{our} for the analysis of the experimental scheme in fluorescence detected 2D spectroscopy. In the conventional photon echo 2D spectroscopy 3 noncollinear laser pulses generate coherent signal in phase matched directions. The signal is mixed with the 4th one, so called local oscillator pulse, which allows phase sensitive field detection~\cite{J2003, HZ2011}. Analysis of the signal coming out with a given combination of $k$'s for various choices of the time delays between the pulses gives information of quantum dynamics and dissipation of the system. In the modulation spectroscopy approach, all four beams are collinear, but the base frequencies are modulated by the acoustic frequencies $\phi_1$, $\phi_2$, $\phi_3$, $\phi_4$. As the result, the signal (fluorescence or photocurrent) is expected to contain harmonics at frequencies ${\boldsymbol m}\cdot{\boldsymbol \phi}=m_1\phi_1+m_2\phi_2+m_3\phi_3+m_4\phi_4$, ($m_i$ are integers, positive or negative, subject to the constrain $m_1+m_2+m_3+m_4=0$). The corresponding relative signal amplitudes should give information of the internal parameters of the system. The spatial separation of signals in the original 2D scheme is replaced by filtering out the proper Fourier component of the signal.  A special attention in our research is paid to the description of the accumulation effects, i.e. the case of a system with a long life-time excited state. 

Our analysis is based on solution of a system of Bloch equations for an atom driven by a train of phase-modulated light pulses. Contrary to Ref.~\cite{our}, where the two-photon absorption process has been involved, we consider a near-resonance processes. As we demonstrate below, the theoretical estimation of fluorescence signal contains the sought harmonics. We estimate their amplitudes and corrections to the amplitudes appearing due to the accumulation effects in the case of long-living excitation and analyse nontrivial features in the structure of the obtained harmonics.   

\begin{figure}
\includegraphics[height=2.4cm,width=8.5cm]{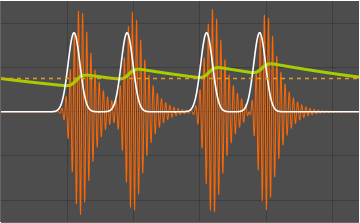}\vspace{4pt}\\
\includegraphics[height=3.cm,width=8.4cm]{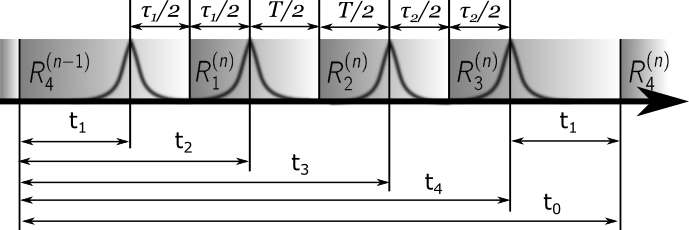}
\caption{\small {\it Upper plot}: Behaviour of $X$ (orange) and $Z+1/2$ (green) components of the Bloch vector for a single quad-pulse in the regime of non-trivial steady-state solution, when the system starts from partually excited state and returnes back to the same state. The decay parameters are chosen in the way that the coherence (component $X$) efficiently decays between the pulses, while the fluorescence mainly generated during the long time-intervals $t_1$.  The pulses envelops are plotted in white. {\it Bottom}: Scheme of notations for the time intervals used in the article. The actual relation between time scales, $t_0\approx 2 t_1\gg \tau_1\sim T\sim\tau_2$, are not shown on the figure. \label{Fig1}}
\end{figure}

{\bf Model of two-level system driven by frequency modulated pulsed field.} The Hamiltonian of two-level system interacting with light can be written in terms of Pauli matrices~\cite{Pauli}, it has the form $\mathcal{H}=\frac{\hbar}{2}\omega_0\sigma_3-\hbar\Omega(t)\sigma_1$. The $\omega_0$ is the transition energy and $\Omega(t)$ describes interaction with the time-dependant external field, 
\begin{equation}\label{Om}
\Omega(t)=\frac{V}{\Delta}\sum_{n=-\infty}^\infty\sum_{i=1,2,3,4}\Theta_\Delta(t-t_i-t_0 n)\cos (\omega_i t).
\end{equation}
Here $n$ counts the quad-pulses (we use the notion quad-pulse as a single word for the neighbouring four pulses, see figure~\ref{Fig1}), while summation over $i$ runs over four single pulses within the single quad-pulse, $\omega_i=\omega+\phi_i$ and $V$ is the coupling of the system with the field. The function $\Theta_\Delta(t)$ describes the envelope of the single pulse with the pulse duration $\Delta$.

The standard theoretical approach for prediction of 2D spectra is based on the time-dependant perturbation theory~\cite{Mbook} developed for the solution of density matrix evolution equation and is a useful tool for bookkeeping of diagrams representing various quantum pathways of light-matter interaction. This theory, while being very efficient in for calculation of the response functions in spectroscopy, is hard to be used for derivation of the solution describing the accumulation effects. To this end, the language of Bloch equation is more convenient. In this language various diagrams are counted automatically by multiplication of the evolution matrices. Nevertheless, the language of ladder diagrams is used on the figure~\ref{Fig4}.

 Evolution of the Bloch vector $R$ with components $X$, $Y$ and $Z$ obeys the matrix equation~\cite{AEbook} 
\begin{equation}\label{Bl}
\frac{dR}{dt}= M R-\frac{\Gamma}{2}\left(\begin{array}{c}
 0\\
0\\
1
\end{array}\right),\quad
 M=\left(\begin{array}{ccc}
 -\gamma&-\omega_0&0\\
 \omega_0&-\gamma&2\Omega\\
 0&-2\Omega&-\Gamma
\end{array}\right).
\end{equation}
Below for the constant vector in~(\ref{Bl}) we use the notation $\Xi$, i.e. $\Xi=\left( 0,0,1\right)^T$. Remind, that the components of the vector $R=(X,Y,Z)^T$ are connected with the components of the density matrix, $\rho$, such that~\cite{Pauli}  $X=\frac{1}{2}\tr \rho\sigma_1$, $Y=\frac{1}{2}\tr \rho\sigma_2$, and $Z=\frac{1}{2}\tr \rho\sigma_3$. The component $Z$ is equal to the difference of the diagonal elements, $(\rho_{ee}-\rho_{gg})/2$, and, since $\tr\rho=1$, the value $Z=-1/2$ corresponds to the system in the ground state ($\rho_{gg}=1$), while the $Z+\frac{1}{2}$ is the probability to find system in the excited state. The constants $\gamma$ and $\Gamma$ describe decay rate of coherence (the off-diagonal elements of the density matrix) and of the excited state due to fluorescence and the non-radiative processes, respectively. In the physically relevant picture $\gamma$ is large, while the other parameter, $\Gamma$, is small.  
The relation between these two parameters means that the fluorescence is generated mainly between the quad-pulses and is proportional to the exponent $e^{-2\Gamma t_1}$, while the decoherence effects are essential on the time scales comparable with the time intervals between the pulses, see figure~\ref{Fig1} top. 

In our model we employ the usual approximations for the 2D spectra analysis. We assume the time ordering of pulses and semi-impulsive limit~\cite{Mbook}. It means that the pulses are well separated and short compared to any time scale (including the ones corresponding to the detuning $\epsilon=\omega_0-\omega$ and the coupling $V$ of the system with the field), but long compared to the oscillation period of the light field, $\omega^{-1},\omega_0^{-1}\ll \Delta$. Notations for the time-intervals used in the text are shown on the figure~\ref{Fig1} bottom. The time-scales are ordered in the following way:
{\small \begin{equation}
\gamma^{-1}\sim\tau_1\sim T\sim\tau_2\nonumber \ll\Gamma^{-1}\sim t_1\sim t_0\ll \frac{2\pi}{\phi_0}\sim\frac{2\pi}{\phi_1}\sim\frac{2\pi}{\phi_2}\sim\frac{2\pi}{\phi_3}.
\end{equation}}
In experiments often the time interval $T$ is much longer then the intervals $\tau_1$, $\tau_2$, but this fact do not influence on our further consideration, as soon as we assume that the fluorescence during interval $T$ is negligible.

The solution of the equation~(\ref{Bl}) at time $t$, for the given initial condition $R(t_{\mathrm{init}})$ taken at time $t_{\mathrm{init}}$, can be generically represented in the form~\cite{PY1990}
\begin{equation}\label{sol}
R(t)=U(t,t_{\mathrm{init}})\left[R(t_{\mathrm{init}})-\frac{\Gamma}{2}\int_{t_{\mathrm{init}}}^tU^{-1}(t',t_{\mathrm{init}})\Xi dt'\right ],
\end{equation}
where $U(t,t_{\mathrm{init}})$ is an orthogonal 3$\times$3 matrix, which satisfies the homogeneous matrix equation $\frac{d}{dt}U= M U$, subject to some initial condition $U(t_{\mathrm{init}},t_{\mathrm{init}})$. In case of zero field, $\Omega(t)=0$, the matrix $U(t,t_{\mathrm{init}})$ is a matrix of free evolution of the system, $U_0(t-t_{\mathrm{init}})$, see equation~(\ref{U0}) in the appendix. Since the pulses are well separated, the electric field of light between the pulses is zero, $\Omega(t_{\mathrm{init}})=0$, so that $U(t_{\mathrm{init}},t_{\mathrm{init}})=\1$.  Therefore, one can seek for solution around each single pulse and glue them up. The whole calculation scheme can be presented in the following way
\begin{eqnarray}\label{R1}
R_1^{(n)}&=&U^{(n)}_1 R_4^{(n-1)}+ F^{(n)}_1,\\
R_2^{(n)}&=&U^{(n)}_2 R_1^{(n)}+ F^{(n)}_2,\\
R_3^{(n)}&=&U^{(n)}_3 R_2^{(n)}+ F^{(n)}_3,\\
R_4^{(n)}&=&U^{(n)}_4 R_3^{(n)}+ F^{(n)}_4.\label{R4}
\end{eqnarray}
The vectors $R_i^{(n)}$ are the vectors obtained after evolution around a single pulse, $R_i^{(n)}=R\big((n-1)t_0+\tilde t_i\big)$ ($\tilde t_i$ is the time after the $i$th pulse where $\Omega(\tilde t_i)=0$, i.e. $\tilde t_1=t_1+\tau_1/2$, $\tilde t_2=t_2+T/2$, $\tilde t_3=t_3+\tau_2/2$, and $\tilde t_4=t_0$), which, in turn, serves as initial vector for the next stage of evolution. The evolution matrix, $U((n-1)t_0+t',(n-1)t_0+\tilde t_{i-1})$ calculated in the local time frame, $\tilde t_{i-1}<t'<\tilde t_i$, generate matrices $U^{(n)}_i\equiv U((n-1)t_0+\tilde t_i,(n-1)t_0+\tilde t_{i-1})$. Note that the $n$-dependence of $U^{(n)}_i$ arises due to the global time dependence of the excitation field phase in $\Omega(t)$, eq.~(\ref{Om}). The vector $F^{(n)}_i$ originates from the second term in the solution~(\ref{sol}). Within the above assumptions the problem can be solved analytically. Explicit forms of $U^{(n)}_i$ and  $F^{(n)}_i$, and comparison with numerics are given in appendix, see equations~(\ref{Un}) and~(\ref{Fn}) and the text around. 

For the purposes of the present investigation we are interested in certain Fourier components of the fluorescent signal. Such analysis can be done by analogy with the one provided in Ref.~\cite{our}. Fluorescence is proportional to the population of the excited state, $Z(t)+1/2$. As soon as we are not interested in the way the system was prepared the Fourier transform of the fluorescent signal, $PL(\nu)$, can be presented as a series of Fourier integrals calculated in the vicinity of each quad-pulse, such that the initial instant is in $-\infty$:
\begin{multline}\label{PL0}
PL(\nu)=\int_{-\infty}^{+\infty}\left(Z(t)+\frac{1}{2}\right)e^{-\i\nu t}dt\\
=\sum_{n=-\infty}^{+\infty} e^{-\i\nu t_0 n} \int_0^{t_0}\left(Z\big(t'+t_0 n\big)+\frac{1}{2}\right)e^{-\i\nu t'}dt'
\end{multline}
The main contribution to the fluorescence comes form the longest time interval $2t_1$, where the population decays exponentially with the rate $\Gamma$, namely the time behaviour of $Z(t)$ in this time interval is given by $\left(Z_4^{(n)}+\frac{1}{2}\right)e^{-\Gamma t}$, where $Z_4^{(n)}$ is $Z$-component of the vector $R_4^{(n)}$ calculated from equation~(\ref{R4}). The function  $PL(\nu)$ after this assumption can be approximated as 
\begin{multline}\label{PL}
PL(\nu)\approx\sum_{n=-\infty}^{+\infty} e^{-\i\nu t_0 n}\int_0^{2t_1\approx t_0}\left(Z_4^{(n)}+\frac{1}{2}\right)e^{-\Gamma t-\i\nu t}dt\\
\approx\frac{1}{\Gamma+\i\nu}\sum_{n=-\infty}^{+\infty} e^{-\i\nu t_0 n} \\\times \left(\left(Z_4^{(n)}+\frac{1}{2}\right) -\left(Z_4^{(n-1)}+\frac{1}{2}\right) e^{-2\Gamma t_1}\right)
\end{multline}
Therefore, $Z^{(n)}_4$ is the only component of the solution we are interested in. The form of expression~(\ref{PL}) for the fluorescent signal reproduces the one obtained in ref.~\cite{our} for the case of linear kinetics. Thus, the result~(\ref{PL}) has a universal nature and reflects the exponential decay of the excited state. 

{\bf Calculation of the fluorescent signal in case of long-living excited state.} 
In our analysis we assume that (i) $t_1\gg 1/\gamma$ and the quantum coherence does not contribute to the formation of the accumulated state; (ii) the oscillations attributed by the frequencies $\phi_i$ are slow, $\phi_i t_0\ll 1$, so that the nearest quad-pulses generate identical excitations of the system; (iii) the excited state life-time $1/\Gamma$ is comparable with $t_0\simeq 2t_1$, but less then $2t_0$, so that $e^{-\Gamma t_1}$ is a small parameter of the model. Therefore, to find the accumulated state solution it is enough to require equivalence of $Z$ components of the Bloch vector before and after the quad-pulse, $Z_4^{(n-1)}=Z_4^{(n)}$. Indeed, $X$ and $Y$ components decay faster then the time interval between the quad-pulses, $t_1$, so in the initial condition for the second quad-pulse they can be put to zero and the initial vector $R_4^{(n-1)}$ is proportional to $\Xi$. Population of the accumulated state is the same before and after the quad-pulse, see figure~\ref{Fig1}. 

From the set of vector equations~(\ref{R1})-(\ref{R4}), which becomes closed by setting $R_4^{(n-1)}= Z_4^{(n)}\Xi$, one can find the solution for $Z_{4\,\mathrm{steady}}^{(n)}$. To do this we solve  the system~(\ref{R1})-(\ref{R4}) perturbatively. The solution is an expansion over the small factor $e^{-\Gamma t_1}$:
\begin{equation}\label{Z4st}
Z_{4\,\mathrm{steady}}^{(n)}=-\frac{1}{2}+Z_{\mathrm{pr}}^{(n)}e^{-\Gamma t_1}\left(1-Z_{\mathrm{nu}}^{(n)} e^{-2\Gamma t_1}\right)+\mathcal{O}(e^{-5\Gamma t_1}).
\end{equation}
The primary contribution $Z_{\mathrm{pr}}^{(n)}$ is the one obtained from interaction of the quad-pulse with two-level system, which initially is in the ground state. Formally,  $-\frac{1}{2}+e^{-\Gamma t_1} Z_{\mathrm{pr}}^{(n)}$ is $Z$ component of $R_4^{(n)}$ obtained from~(\ref{R1})-(\ref{R4}) at $R_4^{(n-1)}$ set to $-\frac{1}{2}\Xi$. The second term in~(\ref{Z4st}) describes ''interference`` of the newly generated excitation with the primary state. The part $Z_{\mathrm{nu}}^{(n)}$ can be found from the solution of the problem~(\ref{R1})-(\ref{R4}) with $R_4^{(n-1)}= z\,\Xi$. Then $Z$ component of  $R_4^{(n)}$ becomes $Z_4^{(n)}=-\frac{1}{2}+Z_{\mathrm{pr}}^{(n)}e^{-\Gamma t_1}+\left(z+\frac{1}{2}\right) Z_{\mathrm{nu}}^{(n)} e^{-2\Gamma t_1}$. Setting $z=-\frac{1}{2}$ we return back to the primary contribution. To find the accumulated state one has to take $z$ to be again equal to $Z_4^{(n)}$ and solve the resulting linear equation. Keeping the lower order terms, up to the $e^{-3\Gamma t_1}$, we arrive to the formula~(\ref{Z4st}). Note, that the higher order terms, which are denoted by $\mathcal{O}(e^{-5\Gamma t_1})$ come from accounting of three and more quad-pulses. The explicit forms of $Z_{\mathrm{pr}}^{(n)}$ and $Z_{\mathrm{nu}}^{(n)}$ were derived by method of functional programming, their expressions can be obtained by the formulas~(\ref{Zpr}) and~(\ref{Znu}) and the tables given in the appendix. 

Having calculated the $Z$ component of the accumulated state one can find the fluorescence by the formula~(\ref{PL}). The laser pulse leads the system through the coherence to the fully excited state, in which the fluorescent signal is generated. After interaction with the $i$th pulse the components of Bloch vector get a term with the phase factor $e^{\i m_i\phi_it_0(n-1)}$, since the $i$th carrier frequency from the $n$th quad-pulse has a phase shift $\phi_it_0(n-1)$. The value of $m_i$ is determined by the effective number of interactions with the $i$th pulse in the primary contribution and by the sum of the effective numbers of interactions from the nearest quad-pulses for the accumulation part. Eventually $Z$ component contains a sum of terms distinguished by the phases  $e^{\i (\boldsymbol m\cdot \boldsymbol \phi) t_0(n-1)}$. Summation in the formula~(\ref{PL}) can be easily implemented using the formula $\frac{1}{2\pi}\sum_{n=-\infty}^\infty e^{\i n \alpha}=\lim_{N\to\infty}\frac{\sin(N+1/2)\alpha}{2\pi\sin \alpha/2}=\delta(\alpha)$. The resulting expression for the fluorescence after substitution of~(\ref{Z4st}) into~(\ref{PL}) has the form
\begin{multline}\label{PL1}
PL(\nu)\approx\frac{1- e^{-2\Gamma t_1}}{\Gamma+\i\nu}\sum_n e^{-\i\nu t_0 n}Z_{pr}^{(n)} \left(1-Z_{nu}^{(n)} e^{-2\Gamma t_1}\right)\\ =\sum_{\boldsymbol m}\delta(\boldsymbol m\cdot\boldsymbol \phi-\nu) A_{\boldsymbol m}, 
\end{multline}
with the amplitudes
\begin{equation}
A_{\boldsymbol m}=\frac{(1- e^{-2\Gamma t_1})\sin^2V}{4(\Gamma+\i\nu)t_0}\left( \mathcal{B}_{\boldsymbol m}+e^{-2\Gamma t_1}\sin^2V\delta \mathcal{B}_{\boldsymbol m}\right).
\end{equation}
The sum in~(\ref{PL1}) runs over the integer-valued vectors $\bm m=(m_1,m_2,m_3,m_4) $ subject to the constraint $m_1+m_2+m_3+m_4=0$.The correction $\delta\mathcal{B}_{\boldsymbol m}$ can be obtained by the formula
$
\sin^2V\delta\mathcal{B}_{\boldsymbol m}=\sum_{\tilde{\boldsymbol m},\tilde{\boldsymbol m}'}\mathcal{B}_{\tilde{\boldsymbol m}}\left(\mathcal{B}_{\tilde{\boldsymbol m}'}+\mathcal{C} _{\tilde{\boldsymbol m}'}\right)\delta_{\tilde{\boldsymbol m} +\tilde{\boldsymbol m}'-\boldsymbol m},
$ here $\delta_{\boldsymbol m}$ is the Kronecker symbol. Explicit formulas for the complex-valued coefficients $\mathcal{B}_{\boldsymbol m}$ and $\mathcal{C} _{\boldsymbol m}$ are given in the appendix. 
\begin{figure}[t]
\includegraphics[scale=0.4]{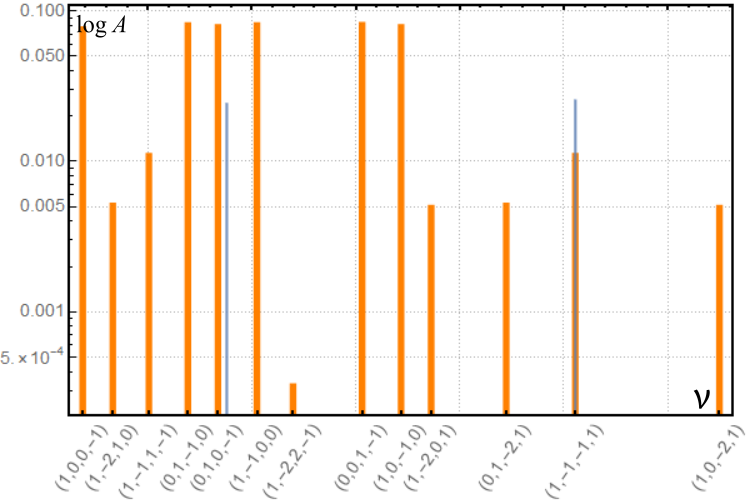}\\
\caption{\small A typical picture of the primary peaks and the absolute values of their amplitudes (in logarithmic scale). The data for this realisation were calculated for incommensurated values of frequencies $\phi_i$ ordered in the way $\phi_1<\phi_4<\phi_2<\phi_3$. Particular choice of parameters was made, in such a way to separate the peaks. The orange peaks are marked by their indexes $\boldsymbol m=(m_1,m_2,m_3,m_4)$, thus each peak is positioned at $\nu=m_1\phi_1+m_2 \phi_2+m_3\phi_3+m_4\phi_4$. The minimal order of light-matter interaction defines the peaks' amplitudes, it is determined by the quantity $\abs{m_1}+\abs{m_2}+\abs{m_3}+\abs{m_4}$. In addition the peaks' amlitudes include the exponential factors depending on intervals $\tau_1$, $T$ and $\tau_2$. The blue thin lines are corrections from the non-trivial steady-state, the plotted values of corrections are taken without the factor $e^{-2\Gamma t_1}$. Only two of many corrections are plotted: the right blue line is the correction to the peak $(1,-1,-1,1)$ and the left blue line corresponds to the newly generated peak $(1,1,-1,-1)$, see eqns.~(\ref{A11}) and~(\ref{A22}). \label{Fig2}}
\end{figure}

{\bf Positions of the peaks, amplitudes and corrections.} Each $\delta$-function in~(\ref{PL1}) is responsible for one peak on the Fourier transform plot, figure~\ref{Fig2}. Within the assumptions made in the model, there are 13 primary peaks (orange lines on the figure) generated by each quad-pulse and 49 secondary peaks generated by the accumulated steady-state. This numbers do not include the peaks appearing due to the symmetry of the coefficients $\mathcal{B}_{-\boldsymbol m}=\mathcal{B}^*_{\boldsymbol m}$, $\delta\mathcal{B}_{-\boldsymbol m}=\delta\mathcal{B}^*_{\boldsymbol m}$. These peaks are positioned on the negative half-line of the $\nu$ axis symmetrically to the ones plotted on the picture (figure~\ref{Fig2}) and have complex conjugated amplitudes. One more peak appears at the origin, $\nu=0$, and is not shown.
 
Positions of the peaks in the frequency domain $\nu$ (see eq.~(\ref{PL1})) form an unstructured set, see figure~\ref{Fig2}. There is no regular ordering of the peaks. To find the indexing vector $\bm m$ one has to calculate the values $(\boldsymbol m\cdot\boldsymbol\phi)$ and find the appropriate value of $\nu$. The strength of amplitudes is mainly determined by the order of interaction with the quad-pulse, $V^2$, $V^4$ or $V^6$. The minimal order of interaction can be calculated as the sum $\abs{m_1}+\abs{m_2}+\abs{m_3}+\abs{m_4}$. The peak amplitudes also include exponential factors $e^{-\gamma(\tilde t_i-\tilde t_{i-1})}$ and $e^{-\Gamma(\tilde t_i-\tilde t_{i-1})}$, i.e. they depend on the time intervals between the pulses. 

\begin{figure}[b]
\includegraphics[width=\columnwidth]{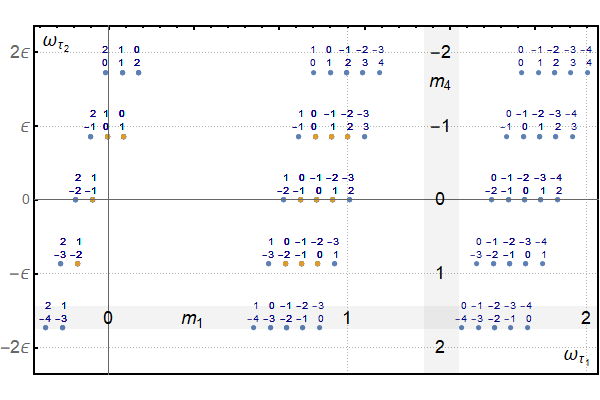}\vspace{-11pt}\\
\includegraphics[width=\columnwidth]{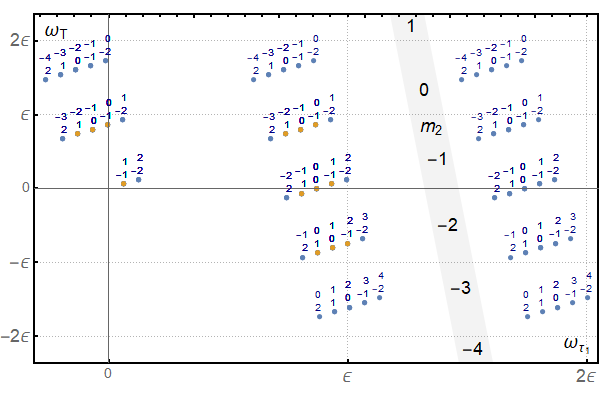}
\caption{\small Positions of primary (orange) and secondary (blue) peaks in the coordinates $\omega_{\tau_1}$ v.s. $\omega_{\tau_2}$ on the upper and $\omega_{\tau_1}$ v.s. $\omega_{T}$ on the bottom plots. Values of two indexes from the corresponding vector $\boldsymbol m$ can be read off from the coordinates in the grey bands, the other two indeces  are given close to the point itself in the order $\atop{m_2}{m_3}$ on the upper and $\atop{m_3}{m_4}$ on the bottom plots. The parameters were kept at the same value as for simulations in~figure~\ref{Fig2}. See text for further explanations.\label{Fig3}}
\end{figure}The set of the peaks becomes structured in another representation, which is obtained after taking the Fourier transformation of the expression~(\ref{PL1}) for the fluorescence over the parameters $\tau_1$ and $\tau_2$ (or $T$). In coordinates $(\omega_{\tau_1},\omega_{\tau_2})$ (or $(\omega_{\tau_1},\omega_{T})$) it forms some regular  lattice-like structure, see figure~\ref{Fig3}. The peaks are positioned around the multiples of the detuning frequency $\epsilon$ and arranged with respect to their indices $\boldsymbol m$. The shift from the exact values of the detuning is due to the factor $\boldsymbol m\cdot\boldsymbol \phi$ coming additively to the $\epsilon$. The primary peaks (orange dots on the plot, fig~\ref{Fig2}) are placed around six points $\omega_{\tau_1},\abs{\omega_{\tau_2}}, \abs{\omega_{T}}\lesssim \epsilon$, while the secondary peaks can have positions around $\pm 2\epsilon$. The latter peaks is a results of effective multiplication of similar or identical interaction processes taken place in the neighbouring quad-pulses. 

Two important remarks has to be done here. Presentation in coordinates $(\omega_{\tau_1},\omega_{\tau_2})$ allows us to group the signals systematically. While the detuning frequency $\epsilon=\omega-\omega_0$ has a clear definition, in case of short spectrally broad laser pulses, it cannot be experimentally determined. 
Therefore the plots on the figure~\ref{Fig3} is a convenient way to represent all peaks at once, but not the real picture that can be observed in experiment. The second remark, is that keeping of the constant phase differences in the long series of quad-pulses is important for observation of harmonics in the fluorescent signal. Randomisation of the phase would lead to destruction of the obtained picture. Indeed, the summation described in the text above the equation~(\ref{PL1}) with any randomisation of phases would result in the sum $\frac{1}{2\pi}\sum_{n=-N}^N e^{\i \big((\boldsymbol m\cdot\boldsymbol \phi)-\nu\big)t_0 n+\i \alpha_n}$ with $\alpha_n$ randomly chosen for each $n$. In the limiting case of statistically independent random $\alpha_n$ distributed uniformly from $-\pi$ to $\pi$ the Fourier transform would result in one peak at $\nu=0$ as $N$ goes to infinity, while in the intermediate cases for distribution of $\alpha_n$ one expects~\cite{AHN2002} some distribution of peak positions around their canonical values given by $(\boldsymbol m\cdot\boldsymbol \phi)$, which can be sensitive to the particular choice of $N$.

Compare now the amplitudes of primary peaks and corrections to them from accumulated state. As an example, we consider the most interesting peak ${\boldsymbol m}=(1,-1,-1,1)$ and the newly formed peak ${\boldsymbol m}=(1,1,-1,-1)$.  The exact expressions for the corrections are given in the appendix, eqs.~(\ref{A1}),~(\ref{A2}). It is instructive to compare the peak amplitude in the leading orders over $V$ and $e^{-2\Gamma t_1}$. From the eqs.~(\ref{A1}),~(\ref{A2}) we obtain
{\small \begin{eqnarray}
A_{1,-1,-1,1}&\simeq &\frac{V^4}{8\Gamma t_0}e^{-\gamma(\tau_1+\tau_2)}\left(1+\frac{1}{4}e^{-2\Gamma t_1}\right),\label{A11}\\
A_{1,1,-1,-1}&\simeq &\frac{V^4}{4\Gamma t_0}e^{-2\Gamma t_1-\gamma(\tau_1+2T+\tau_2)}.\label{A22}
\end{eqnarray}}
As one can see the correction to $A_{1,-1,-1,1}$, the last term in the brackets in eq.~(\ref{A11}), as well as the amplitude $A_{1,1,-1,-1}$ has the same order of magnitude as the amplitude of the primary peak and only the factor $e^{-\Gamma t_1}$ can significantly suppress the above corrections. The ratio of the amplitudes $A_{1,1,-1,-1}/A_{1,-1,-1,1}\approx 2 e^{-2\Gamma t_1-2\gamma T}$ gives us an access to the parameter $\gamma$. Note, that the combination of phases $\phi_1+\phi_2-\phi_3-\phi_4$ should correspond to the signal attributed in the literature as the double quantum coherence signal~\cite{MOY2007}, which can appear in the three or more level atom due to formation of the coherence between the ground state and the upper excited level. In present theory the signal is the result of accumulation. Namely, the signal amplitude is a product of two amplitudes with the indexes $(1,0,0,-1)$ and $(0,1,-1,0)$. Since they both have the same order of $V^2$ this explains the strength of the total signal proportional to $V^4$. Some more explanations regarding the structure of  eqs.~(\ref{A11}) and~(\ref{A22}) are given in the caption of the figure~\ref{Fig4}.
\begin{figure}[b]
\parbox[b]{2cm}{\small $A'_{1,-1,-1,1}\simeq$\vspace{17pt}}\includegraphics[scale=0.35]{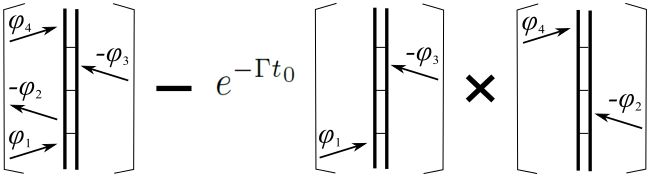}\\
\parbox[b]{2cm}{\small $A'_{1,1,-1,-1}\simeq$\vspace{17pt}}\includegraphics[scale=0.35]{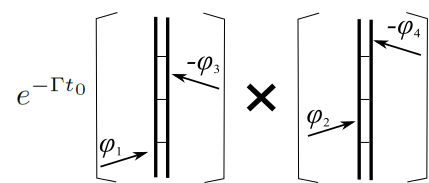}
\caption{\small Schematic representation of some contributions into the peak amplitudes $A_{1,-1,-1,1}$ and $A_{1,1,-1,-1}$ in terms of ladder diagrams~\cite{Mbook}, see eqs.~(\ref{A11}), and~(\ref{A22}) and also eq.~(\ref{Z4st}) and the text around it. The fluorescence signal is detected from fully excited atoms, therefore each ladder diagram with necessety contain the pathway from ground to the excited state. The  primary contribution to the signal $(1,-1,-1,1)$ results from interaction with the single quad-pulse, while correction is a product of the factors coming from the two-legs diagrams due to accumulation of th eexcitatioon from the previous quad-pulse, the effect of accumulation is suppressed by the exponent $e^{-2\Gamma t_1}$ and since $2t_1\approx t_0$ it is replaced by $e^{-\Gamma t_0}$ on the picture. One of the diagram describes excitation of the atom from the ground state by the second quad-pulse, while the first diagram comes with the minus sign and describes change of the ground state population out of specific excitation of the atom by the first quad-pulse. The other contributions to the peak amplitudes can be obtained by another suitable choice of the legs combinations.\label{Fig4}}
\end{figure}

{\bf Conclusion.} In this work we studied the model of two-level system driven by light coming in quad-pulses with modulated frequencies. Our analysis was based on analytic solution of the system of the optical Bloch equations under standard assumptions used in analysis of spectroscopic problems. It was shown that the Fourier transform of the fluorescent signal contains a number of oscillating modes with the frequencies ${\boldsymbol m}\cdot {\boldsymbol \phi}$. Amplitudes of the primary peaks, originated from each single quad-pulse, get corrections when one takes into account the accumulation effects. Moreover, these effects lead to formation of new (secondary) peaks. The corrections as well as amplitudes of the secondary peaks are of the same order of magnitude as the amplitudes of the primary peaks, the only suppressing factor is the exponential decay of the excited state from one quad-pulse to the other. Therefore, the cumulative effects have to be taken into account in the analysis of phase-modulated harmonic light spectroscopy data. Note, also that the ratio of the amplitudes of primary and secondary peaks can be used for revealing of the internal parameters of the system, see for instance eqs.~(\ref{A11}) and~(\ref{A22}). 

There are other effects that can change the peaks' amplitudes, which have not been considered above. First, our consideration is based on certain physical assumptions, such as semi-impulsive limit and perfect time ordering of pulses. Accounting of effects beyond these simplifications can bring additional corrections into the amplitude values. More important, however, is the following observation. The peaks obtained in our derivation, based on Bloch equation, should be considered as a two-level system response on light excitation averaged over ensemble of many identical systems. In the work Ref.~\cite{M2016} it is noticed that generation of higher harmonics in the signal can be caused by many-particle effects. In fact, the same many-particle arguments lay in the basis of our classical consideration of the modulation spectroscopy in Ref.~\cite{our}. Therefore, the corrections to the primary peaks can have various sources. This part of the problem certainly requires further theoretical efforts to formulate a complete physical picture of the interplay between the accumulation and many-particle phenomena.
\vspace{10pt}




{\bf Acknowledgements.} Authors thanks Khadga J. Karki and Shaul Mukamel for useful discussions.

\newpage
\section*{Appendix A}

\begin{widetext}

Explicit forms of matrices used in the paper. Matrix of free evolution of the system has the form
\begin{equation}\label{U0}
U_0(t)=\left(\begin{array}{ccc}
e^{-\gamma t}\cos\omega_0 t&-e^{-\gamma t}\sin\omega_0 t&0\\
e^{-\gamma t}\sin\omega_0 t&e^{-\gamma t}\cos\omega_0 t&0\\
0&0&e^{-\Gamma t}
\end{array}\right).
\end{equation}
It is solution of the equation $\frac{d U_0(t)}{dt}=\tilde M U_0(t)$, where $\tilde M$ is matrix $M$ from eq.~(\ref{Bl}) with $\Omega=0$. To find matrix $U(t)$ we make the substitution $U(t)=U_0(t)U_I(t)$ and consider the problem in interaction picture~\cite{PY1990}
\begin{equation}\label{Bl1}
\frac{dU_I(t)}{dt}= 2\Omega(t)U_0^{-1}(t)\left(\begin{array}{ccc}
 0&0&0\\
 0&0&1\\
 0&-1&0
\end{array}\right)U_0(t)U_I(t).
\end{equation}
Under assumptions made in the paragraph around the equation~(\ref{sol}) the matrix equation~(\ref{Bl1}) can be solved approximately in the vicinity of each pulse to obtain expressions for $U^{(n)}_i$ and  $F^{(n)}_i$. They are
\begin{multline}\label{Un}
U^{(n)}_i=U_0(\tilde t_i-\tilde t_{i-1})\\\times\left[\1+\left(\begin{array}{ccc}
-2\sin^2 f_i\sin^2 \frac{V}{2}&- \sin 2 f_i\sin^2\frac{V}{2}& e^{-(\Gamma-\gamma )(t_i-\tilde t_{i-1})}\sin f_i\sin V\\
- \sin 2 f_i\sin^2\frac{V}{2}&-2\cos^2 f_i\sin^2\frac{V }{2}& e^{-(\Gamma-\gamma )(t_i-\tilde t_{i-1})}\cos f_i\sin V\\
-e^{(\Gamma-\gamma )(t_i-\tilde t_{i-1})}\sin f_i\sin V&-e^{(\Gamma-\gamma )(t_i-\tilde t_{i-1})}\cos f_i\sin V&-2\sin^2\frac{V}{2}
\end{array}\right)\right];
\end{multline}
\begin{equation}\label{Fn}
F^{(n)}_i=\frac{1}{2}\left(\begin{array}{c}
e^{-\gamma (\tilde t_i-2\tilde t_{i-1}+t_i)}(1-e^{-\Gamma (t_i-\tilde t_{i-1})})\sin V \sin(f_i-\omega_0 (\tilde t_{i}-\tilde t_{i-1}))\\
e^{-\gamma (\tilde t_i-2\tilde t_{i-1}+t_i)}(1-e^{-\Gamma (t_i-\tilde t_{i-1})})\sin V \cos(f_i-\omega_0 (\tilde t_{i}-\tilde t_{i-1}))\\
e^{-\Gamma (\tilde t_i-2\tilde t_{i-1}+t_i)}-1+ (1-e^{\Gamma (\tilde t_i-2\tilde t_{i-1}+t_i)})\cos V
\end{array}\right),
\end{equation}
where we use the notations
$ f_i=\omega_0(t_i-\tilde t_{i-1})-\omega_i t_i-\omega_i t_0(n-1)$.

To check the obtained analytic approximation we compared it with a numerical solution for a single pulse, see figures below.
\begin{center}
a) \includegraphics[scale=0.45]{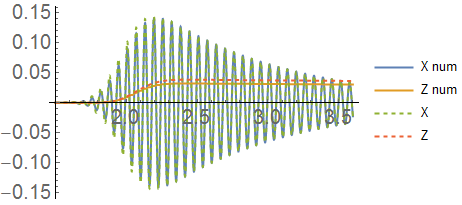}
\hspace{10pt}b) \includegraphics[scale=0.45]{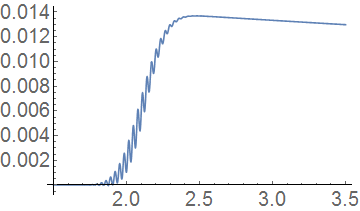}
\end{center}
On the figure a) we compared $X$ and $Z$ components of the numeric solution and of the one obtained within our approximations for Gaussian choice of the pulse envelop, $\Theta_{\Delta=2\sqrt{\pi}\sigma}(t)=e^{-\frac{(t-2)^2}{4\sigma^2}}$ and the frequency $\omega_0\simeq 285 \Delta$, such that there are around six periods of the wave lie within the pulse duration. The detuning $\epsilon=2.82 \Delta$, meaning that system is in resonance. Time dependence of the analytic solution is obtained by replacing $V$ in equations~(\ref{Un}) and~(\ref{Fn}) with $\frac{V}{\Delta}\int_{-\infty}^t\Theta_{\Delta}(t') dt'$.  On the figure b) the value of relative error $\abs{\frac{Z-Z num}{Z}}$ is plotted. The discrepancy reaches not more then 2\%. Note that the further one from the resonance the larger becomes the error.

The non-zero values of coefficients $\mathcal{B}_{\boldsymbol m}$ in eq.~(\ref{PL1}) are given in the table below,
\begin{center}
    \begin{tabular}{ ccc c |c l}
    $m_1$,&  $m_2$,&  $m_3$,&  $m_4$&&$\mathcal{B}_{m_1,m_2,m_3,m_4}$ \\ \hline\hline
    1,& -1,&0,&0 && $
e^{-\Gamma(\tau_2+T)-\gamma \tau_1}e^{\i(\epsilon-\phi_1)\tau_1}\cos^2 V
$
\\     1,& 0,&-1,&0 && $
\frac{1}{2}e^{-\Gamma \tau_2-\gamma(\tau_1+T)}e^{\i(\epsilon-\phi_3)(\tau_1+T)+\i(\phi_1-\phi_3)t_1}(2+\cos V)(3+\cos V)
$
\\
    1,& 0,&0,&-1 && $
\frac{1}{4}e^{-\gamma(\tau_1+T+\tau_2)}e^{\i(\epsilon-\phi_4)(\tau_1+T+\tau_2)+\i(\phi_1-\phi_4)t_1}\left(1+\cos V\right)^2
$
\\
    0,& 1,&-1,&0 && $
-e^{-\Gamma(\tau_1+\tau_2)-\gamma T} e^{\i (\epsilon-\phi_3)T+\i(\phi_1-\phi_3)(t_1+\tau_1)} \left(1-e^{\Gamma \tau_1}-\cos V\right)\cos V
$
 \\
    0,& 1,&0,&-1 && $
-\frac{1}{2}e^{-\Gamma\tau_1-\gamma(T+\tau_2)} e^{\i (\epsilon-\phi_4)(T+\tau_2)+\i(\phi_1-\phi_4)(t_1+\tau_1)} \left(1-e^{\Gamma \tau_1}-\cos V\right)(1+\cos V)
$ \\
    0,& 0,& 1,& -1 && $
e^{-\gamma \tau_2}e^{\i (\epsilon-\phi_4)\tau_2+\i(\phi_1-\phi_4)(t_1+T+\tau_1)}\left(1-e^{-\Gamma T} + e^{-\Gamma T}(1-e^{-\Gamma \tau_1})\cos V+e^{-\Gamma (T+ \tau_1)}\cos^2V\right)
$\\
    1,& -1,&-1,&1 && $
-\frac{1}{2}e^{-\Gamma T-\gamma(\tau_1+\tau_2)} e^{\i (\epsilon-\phi_1-\phi_3+\phi_4)\tau_1 -\i (\epsilon-\phi_4)\tau_2 -\i(\phi_3-\phi_4)(t_1+T)} \sin^2 V
$
\\
    1,& -1,&1,&-1 && $
-\frac{1}{2}e^{ -\Gamma T-\gamma(\tau_1+\tau_2)}e^{\i (\epsilon-\phi_1+\phi_3-\phi_4)\tau_1 +\i (\epsilon-\phi_4)\tau_2 +\i(\phi_3-\phi_4)(t_1+T)}\sin^2 V
$
\\
    1,& -2,&1,&0 && $
-\frac{1}{2}e^{-\Gamma \tau_2-\gamma(\tau_1+T)}e^{\i (\epsilon-2\phi_1+\phi_3)\tau_1-\i (\epsilon-\phi_3)T  -\i(\phi_1-\phi_3)t_1}\left(1-\cos V\right)\cos V
$
\\
    1,& -2,&0,&1 && $
-\frac{1}{4}e^{-\gamma(\tau_1+T+\tau_2)}e^{\i (\epsilon-2\phi_1+\phi_4)\tau_1-\i (\epsilon-\phi_4)(T+\tau_2) -\i(\phi_1-\phi_4)t_1}\sin^2 V
$
 \\
    1,& 0,&-2,&1 && $
-\frac{1}{4}e^{-\gamma(\tau_1+T+\tau_2)}e^{\i (\epsilon-2\phi_3+\phi_4)(\tau_1+T)-\i (\epsilon-\phi_4)\tau_2 +\i(\phi_1-2\phi_3+\phi_4)t_1}\sin^2 V$
\\
    0,& 1,&-2,& 1 && $
\frac{1}{2}e^{-\Gamma\tau_1-\gamma(T+\tau_2)} e^{\i (\epsilon-2\phi_3+\phi_4)T-\i (\epsilon-\phi_4)\tau_2 +\i(\phi_1-2\phi_3+\phi_4)(t_1+\tau_1)} \left(1-e^{\Gamma \tau_1}-\cos V\right)(1-\cos V)
$\\
    1,& -2,&2,&-1 && $
\frac{1}{4}e^{-\gamma(\tau_1+T+\tau_2)}e^{\i (\epsilon+2\phi_3-2\phi_1-\phi_4)\tau_1-\i (\epsilon-2\phi_3+\phi_4)T +\i (\epsilon-\phi_4)\tau_2 -\i(\phi_1-2\phi_3+\phi_4)t_1}\left(1-\cos V\right)^2
$
\\
\hline
    \end{tabular}
\end{center}
Expression for the solution~(\ref{Z4st}) can be obtained from the above table by using the formulas (c.c. means complex conjugation)
\begin{eqnarray}\label{Zpr}
Z_{\mathrm{pr}}^{(n)}&=&\frac{\sin^2 V}{4}\left(\mathcal{B}_{\boldsymbol m}e^{\i(\boldsymbol m\cdot \boldsymbol\phi)(n-1)t_0}+c.c.\right)\nonumber\\&&+\frac{1}{2}\left(1+e^{-\Gamma\tau_2}\cos V+e^{-\Gamma(T+\tau_2)}\cos^2 V+e^{-\Gamma(\tau_1+T+\tau_2)}\cos^3V\right)(1-\cos V),\\
Z_{\mathrm{nu}}^{(n)}&=&-\frac{\sin^2 V}{2}\left((\mathcal{B}_{\boldsymbol m}-\mathcal{C}_{\boldsymbol m})e^{\i(\boldsymbol m\cdot \boldsymbol\phi)(n-1)t_0}+c.c.\right)+e^{-\Gamma(\tau_1+T+\tau_2)}\cos^4V,\label{Znu}
\end{eqnarray}
supplemented by the table of non-zero coefficients $\mathcal{C}_{\boldsymbol m}$:
\begin{center}
    \begin{tabular}{ ccc c |c l}
    $m_1$,&  $m_2$,&  $m_3$,&  $m_4$&&$\mathcal{C}_{m_1,m_2,m_3,m_4}$ \\ \hline\hline
    0,& 1,&-1,&0 && $
e^{-\Gamma\tau_2-\gamma T} e^{\i (\epsilon-\phi_3)T+\i(\phi_1-\phi_3)(t_1+\tau_1)} \left(1-e^{-\Gamma \tau_1}\right)\cos V
$
 \\
    0,& 1,&0,&-1 && $
\frac{1}{2}e^{-\gamma(T+\tau_2)} e^{\i (\epsilon-\phi_4)(T+\tau_2)+\i(\phi_1-\phi_4)(t_1+\tau_1)} \left(1-e^{-\Gamma \tau_1}\right)(1+\cos V)
$ \\
    0,& 0,& 1,& -1 && $
e^{-\gamma \tau_2}e^{\i (\epsilon-\phi_4)\tau_2+\i(\phi_1-\phi_4)(t_1+T+\tau_1)}\left(1-e^{-\Gamma T}+e^{-\Gamma  T}(1-e^{-\Gamma  \tau_1})\cos V\right)
$
\\
    0,& 1,&-2,& 1 && $
-\frac{1}{2}e^{-\gamma(T+\tau_2)} e^{\i (\epsilon-2\phi_3+\phi_4)T-\i (\epsilon-\phi_4)\tau_2 +\i(\phi_1-2\phi_3+\phi_4)(t_1+\tau_1)} \left(1-e^{-\Gamma \tau_1}\right)(1-\cos V)
$
\\
\hline
    \end{tabular}
\end{center}
Explicit form of corrections to the coefficients $\mathcal{B}_{1,-1,-1,1}$ and $\mathcal{B}_{1,1,-1,-1}$ from the non-trivial steady state
\begin{eqnarray}\label{A1}
\delta\mathcal{B}_{1,-1,-1,1}&=&-\frac{1}{4} e^{-T \Gamma -\gamma  (\tau_1+\tau_2)}+\frac{1}{4} e^{-T \Gamma -\gamma  (\tau_1+\tau_2)} \left(1-e^{-\Gamma
\tau_2}\right) \cos V
+\frac{1}{2} e^{-T \Gamma -\Gamma  \tau_2-\gamma (\tau_1+\tau_2)} 
\left(1-e^{-T\Gamma }\right) \cos ^2 V+\nonumber\\
&&\frac{1}{4} e^{-\Gamma \tau_2-\gamma(\tau_1+\tau_2)}\! \left(2e^{-2 \Gamma T }+ e^{-2\gamma T}\right)\!\left(1-e^{-\Gamma\tau_1}\right)\!
\cos^3 V 
+\frac{1}{2} e^{-(\gamma +\Gamma ) (\tau_1+\tau_2)}\left(2e^{-2 \Gamma T }+ e^{-2\gamma T}\right)\! \cos^4\! V,\\\label{A2}
\delta\mathcal{B}_{1,1,-1,-1}&=&\frac{1}{8} e^{-2 \gamma T -\Gamma  \tau_2-\gamma  (\tau_1+\tau_2)} \left(1-e^{-\Gamma  \tau_1}\right) \cos V
+\frac{1}{4}e^{-2 \gamma T -\Gamma  \tau_2-\gamma  (\tau_1+\tau_2)} \cos^2 V+\nonumber\\&&
\frac{1}{8} e^{-2 \gamma T -\Gamma  \tau_2-\gamma  (\tau_1+\tau_2)} \left(1+3 e^{-\Gamma \tau_1}\right)\cos^3 V
+\frac{1}{4} e^{-2 \gamma T -(\gamma
+\Gamma ) (\tau_1+\tau_2)} \cos^4 V.
\end{eqnarray}
\end{widetext}
\bibliographystyle{apsrev4-1} 

\end{document}